%% file: main.tex
\documentclass[12pt]{article}
\usepackage{epsfig,amsmath,calc,rotating,moreverb}
\makeindex
\input{babarsym}

\long\def\inst#1{\par\nobreak\kern 4pt\nobreak {\it #1}\par\vskip 10pt plus 3pt minus 3pt}

\def\acp        {\ensuremath{{\mathcal A}_{CP}}\xspace}

\def\bkstrhoo   {\ensuremath{B^{+} \rightarrow K^{*+}\rho^0}\xspace}
\def\bkstorho   {\ensuremath{B^{+} \rightarrow K^{*0}\rho^+}\xspace}
\def\brhorhoo   {\ensuremath{B^{+} \rightarrow \rho^+\rho^0}\xspace}

\def\Kstar      {\ensuremath{K^*}\xspace}

\def\Kstarz     {\ensuremath{K^{*0}}\xspace}

\def\pipi     {\ensuremath{\pi\pi}}

\def\mes   {\ensuremath{m_{ES}}\xspace}
\def\de   {\ensuremath{\Delta E}\xspace}

\def\dz   {\ensuremath{\Delta z}}

\def\sl   {\ensuremath{ S_{long} }}

\def\piz {\ensuremath{ \pi^0 }}

\def\de      {\ensuremath{\Delta E}\xspace}

\def \Dz {\ensuremath{D^{0}}}
\def \Dstarz {\ensuremath{D^{*0}}}
\def \rhop {\ensuremath{\rho^{+}}}
\def \rhom {\ensuremath{\rho^{-}}}

\begin{document}
\pagestyle{empty}

\begin{flushright}
\babar-CONF-04/34 \\
SLAC-PUB-10641 \\
hep-ex/0408093 \\
August 2004 \\
\end{flushright}

\par\vskip 1.5cm

% Title of the paper
\begin{center}
\Large \bf Measurements of Branching Fraction, Polarization, and Direct-CP-Violating Charge Asymmetry in \boldmath{$\bkstorho$} Decays
\end{center}
\bigskip

\begin{center}
\large The \babar\ Collaboration\\
\mbox{ }\\
\today
\end{center}
\bigskip \bigskip

% Abstract
\begin{center}
{\large \bf Abstract}
\end{center}
\input{abstract}
\vfill
\begin{center}

Submitted to the 32$^{\rm nd}$ International Conference on High-Energy Physics, ICHEP 04,\\
16 August---22 August 2004, Beijing, China

\end{center}

\vspace{0.6cm}
\begin{center}
{\em Stanford Linear Accelerator Center, Stanford University, 
Stanford, CA 94309} \\ \vspace{0.3cm}\hrule\vspace{0.1cm}
Work supported in part by Department of Energy contract DE-AC02-76SF00515.
\end{center}

\newpage

\renewcommand{\thefootnote}{\arabic{footnote}}
\setcounter{footnote}{0}

\input{authors}
\newpage

\pagestyle{plain}

\input{intro}

\input{method}
\input{syst_check}

%%%%%%%%%%%%%%%%%%%%%%%%%%%%%%%%%%%%%%%%%%%%%%%%%%%%%%
%%%%%%%%%%%%%%%%%%%   BIBLIOGRAPHY   %%%%%%%%%%%%%%%%%
%%%%%%%%%%%%%%%%%%%%%%%%%%%%%%%%%%%%%%%%%%%%%%%%%%%%%%
\clearpage
\bibliographystyle{unsrt}
\bibliography{biblio}

%%%%%%%%%%%%%%%%%%%%%%%%%%%%%%%%%%%%%%%%%%%%%%%%%%%%%%
%%%%%%%%%%%%%%%%%%   END OF DOCUMENT   %%%%%%%%%%%%%%%
%%%%%%%%%%%%%%%%%%%%%%%%%%%%%%%%%%%%%%%%%%%%%%%%%%%%%%
\end{document}

%% file: abstract.tex
With a sample of $88.8 \times 10^{6}$ $B\bar B$ pairs produced at PEP-II in 
$e^+ e^-$ annihilation through the $\Upsilon(4S)$ resonance and
recorded with the \babar\ detector, we search for the $\bkstorho$
decay mode. A signal is observed for the first time
with a significance of more than $5\sigma$.
We measure a preliminary branching fraction of
$\BR(\bkstorho) =
      [17.0\pm{}{2.9}\mathrm{(stat)}\pm{}{2.0}\mathrm{(syst)}{}^{+0.0}_{-1.9}(\mbox{non-resonant})]\times 10^{-6}$.
The ``non-resonant'' error corresponds to the uncertainty from non-resonant 
backgrounds not modeled in the fit.
The measurement of the  longitudinal-polarized component to this
vector-vector penguin decay is of special interest. We measure
       $f_L=0.79\pm 0.08\mathrm{(stat)}\pm 0.04 \mathrm{(syst)} \pm 0.02 (\mbox{non-resonant})$. 
We measure the direct-CP-violating charge asymmetry in this mode to be 
$\acp = -0.14 \pm 0.17 \mathrm{(stat)} \pm 0.04 \mathrm{(syst)}$.

%% file: authors.tex
\begin{center}
\small

The \babar\ Collaboration,
\bigskip

%% author list as of 02-Jul-2004 (609 authors)
%
B.~Aubert,
R.~Barate,
D.~Boutigny,
F.~Couderc,
J.-M.~Gaillard,
A.~Hicheur,
Y.~Karyotakis,
J.~P.~Lees,
V.~Tisserand,
A.~Zghiche
\inst{Laboratoire de Physique des Particules, F-74941 Annecy-le-Vieux, France }
% ...

\end{center}\newpage

%% file: intro.tex
\section{INTRODUCTION}

The purpose of this analysis is the simultaneous measurement of the
branching fraction, the longitudinal-polarization component, and the
direct-CP-violating charge asymmetry for the $B\to$ vector-vector
decay mode $\bkstorho$\!, which has previously not been seen
experimentally. 
The measurement of the polarization of this  penguin decay is of 
special interest. There is no tree contribution to the decay amplitude and only 
a very small annihilation process. 
The polarization of the $\bkstorho$ decay mode can be compared to that already
measured for the similar penguin decay $B \to K^* \phi$.
The direct-CP-violating charge asymmetry is defined as:
$$\acp \equiv \frac{N(B^-\rightarrow K^{*0}\rhom) - N(B^+\rightarrow \overline{K}^{*0}\rhop)}{N(B^-\rightarrow K^{*0}\rhom) + N(B^+\rightarrow \overline{K}^{*0}\rhop)}.$$
In this  penguin decay, a value significantly different from zero could be a hint for new physics.

\subsection{Physics motivation}
In the Standard Model, the study of the   penguin decay \bkstorho, combined
with information from other charmless hadronic $B \to$ vector-vector
decays into $K^* \rho$ and $\rho \rho$, allows us to constrain the 
angles $\alpha$ and $\gamma$ of the unitarity triangle, in a way similar to the
study of the $B \rightarrow \pi \pi$ and $B \rightarrow K \pi$ decays
\cite{Fleisher}. The  $\rho \rho$ modes are used to constrain the angle $\alpha$.
The methods that constrain the angle $\gamma$ are based on isospin symmetry relationships 
relating the amplitudes of the different $K^* \rho$ decay modes, as well as on the 
$SU(3)$ flavour symmetry relating the $K^* \rho$ and $\rho \rho$ modes. They also rely on 
the relationship $\alpha = \pi - \beta - \gamma$ and the measured value of the angle 
$\beta$. A constraint on $\gamma$ can already be obtained using only
the charged-\B decays $\bkstorho$, $\bkstrhoo$, and $\brhorhoo$
\cite{NeubertGRL}. The two modes $\bkstrhoo$ and $\brhorhoo$ have been
studied previously \cite{babar_vv,belle_rhorho}.  A  stronger
constraint on $\gamma$ can be obtained by including the $B^0
\rightarrow K^{*+} \rho ^-$ mode, a first study of which
is also being presented at this conference and is based on a similar
analysis technique \cite{Wisconsin}.

Measurements of the rates, polarizations, and direct-CP-violating
asymmetries of these vector-vector decays permit testing
theoretical predictions from the naive factorization~\cite{Ali:1998}
and QCD factorization \cite{Aleksan, kagan_polvv} based models.

In charmless decays of \B\ mesons into two light vector mesons, both
longitudinal and transverse polarization states are possible, but a
large longitudinal-polarization fraction, of order $1-4 \times
\frac{m_{\rho}^2}{m_{B}^2} \approx 0.9$ for $K^* \rho$, is expected
from theory for both tree and penguin decays.
Existing measurements of tree dominated vector-vector charmless modes
$\rho^+\rho^-$ and $\rho^+\rho^0$ show that their longitudinal
component is indeed dominant \cite{babar_vv,belle_rhorho,babar_rhoprhom}.
However, experimental results for the polarization in the pure gluonic
penguin $B^0 \to \phi K^{*0}$ and $B^+ \to \phi K^{*+}$ charmless
processes indicate a transverse polarization fraction of about 0.5
\cite{babar_phikstar, belle_phikstar, babar_vv}.
This difference can be accounted for in the Standard Model 
by increasing the non-factorizable contribution of annihilation diagrams, through the
tuning of some poorly known non-perturbative QCD parameters
\cite{kagan_polvv,ygrossman,Colangelo:2004rd}.
The uncertainty on these parameters makes the accurate {\it a priori}
prediction of the polarization difficult for a given decay mode. However, 
using SU(3) flavour symmetry arguments, the same polarization is expected 
for the \bkstorho decay and for $B \to \phi K^{*}$ decays.  

Thus, the comparison of the polarizations obtained in the two
penguin modes \bkstorho and $\phi K^*$ is of interest since the $\sim 0.5$
transverse polarization fraction observed in $\phi K^*$ decays could also
be due to contributions from new physics in the penguin loop.

Table \ref{tab:summary_predict_meas} summarizes the theoretical predictions from the
naive-factorization model compared to the existing measurements of branching fractions and polarizations 
for the $K^* \rho$ and  $\rho \rho$ decay modes. 
The theoretical predictions for the branching fractions 
are taken from \cite{Ali:1998}. They have been updated in \cite{Gregory:2004}.
The ranges given are obtained by varying the form factors and other parameters 
entering the calculation. The polarizations have also been predicted.  

\begin{small}
\begin{table}[!h]
\caption{Comparison of the predictions from the naive-factorization model to the measurements for the 
branching fractions and polarizations in the $K^* \rho$ and $\rho \rho$ decay modes. The branching fractions are in units of $10^{-6}$. The results for $\bkstorho$ are those reported here.}

\begin{center}
\begin{tabular}{lccccc}
Mode & \BR\ ($10^{-6}$) & \BR\ ($10^{-6}$) & \BR\ ($10^{-6}$)  & Polarization   &  Polarization     \\
    & prediction \cite{Ali:1998} & prediction \cite{Gregory:2004} & Measurement          & prediction \cite{Gregory:2004} &  measurement \\
\hline
\hline
$K^{*+} \rho^0$   & $6.6$ & $6 - 10$ & $10.6 ^{+3.8}_{-3.5}$ & $0.90$ & $0.96 ^{+0.04}_{-0.16}$ \\
$K^{*+} \rho^-$   & $7.0$ & $6 - 10$ & $<17.2$ @ 90\% CL & $0.90$ &  -- \\
$K^{*0} \rho^+$   & $9.0$ & $8 - 12$ & $17.0 ^{+3.5}_{-3.9}$ & $0.90$ & $0.79 \pm {0.09}$ \\
$\rho^{+} \rho^0$ & $6.1$ & $7 - 12$ & $26.4 ^{+6.1}_{-6.4}$ & $0.92$ & $0.89 \pm {0.07}$ \\
$\rho^{+} \rho^-$ & $24.0$ & $20 - 25$  & $27 \pm {9}$ & $0.92$  & $0.98 ^{+0.02}_{-0.09}$ \\
\hline
\hline
\end{tabular}
\end{center}
\label{tab:summary_predict_meas}
\end{table}
\end{small}

\subsection{Angular analysis}

\begin{figure}[!h]\begin{center}
  \includegraphics[width=14cm]{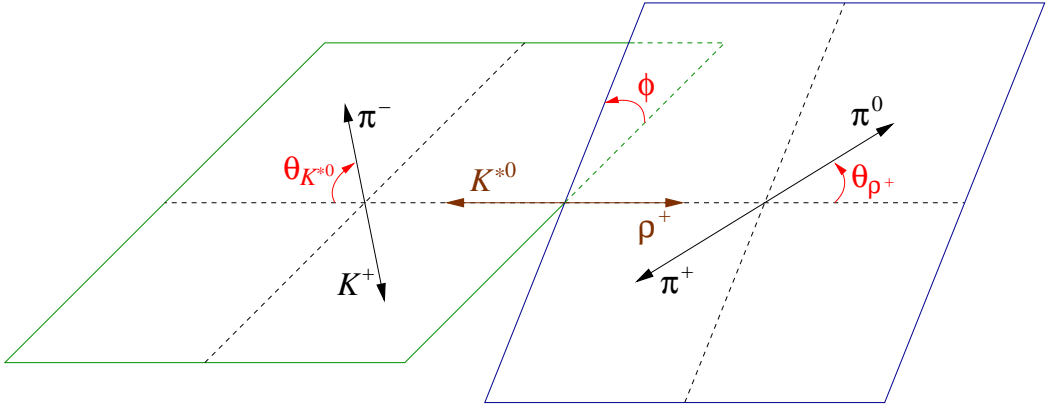}
  \caption{Helicity frames for the vector-vector \bkstorho decay.}
  \label{fig:helicity_dessin}
\end{center}\end{figure}

The analysis is done in the helicity frame
(Fig.~\ref{fig:helicity_dessin}) as a function of the ``helicity
angles'' $\theta_{K^{*0}}$ and $\theta_{\rho^+}$.
The angle $\theta_{K^{*0}}$ ($\theta_{\rho^+}$) is defined as the angle between the
direction of the $K^{*0}$ ($\rho^+$) and the direction of the $\pi^-$
($\pi^0$) coming from its decay in the vector meson's rest frame.
An integration is performed over the angle $\phi$ between the
vector-meson decay planes to simplify our analysis;
this step is straightforward because the detector acceptance is
independent of this angle.
The longitudinal-polarization fraction $f_L$ can be extracted from
the differential decay rate, parametrized as a function of $\theta_{K^{*0}}$
and $\theta_{\rho^+}$ \cite{formal_vvdecay}:
\begin{equation}\label{eqn1}
{1 \over \Gamma}
{ {d^2 \Gamma}\over {d\cos\theta_{K^{*0}} d\cos\theta_{\rho^+}} } \propto
{1\over 4}
(1-f_L)\sin^2\theta_{K^{*0}}\sin^2\theta_{\rho^+}+f_L \cos^2\theta_{K^{*0}} \cos^2\theta_{\rho^+}.
\end{equation}

Experimentally, it is important to measure the branching fraction
and the polarization simultaneously because of their large
correlation.
The two decay products of the vector mesons have in general comparable
momenta when transversely polarized, and asymmetric momenta, with one
high- and one low-momentum decay product, when longitudinally
polarized. As soft particles have lower reconstruction efficiency,
the efficiency for reconstructing longitudinally polarized decays is
about half that for transversely polarized decays.

\section{THE \babar $~$ DETECTOR AND DATASET}
The results presented in this paper are based on data collected in
1999-2002 with the \babar $~$ detector \cite{detector_babar} at the
PEP-II asymmetric $e^+ e^-$ collider of the Stanford Linear
Accelerator Center.  An integrated luminosity of $81.85\,\fb^{-1}$ was recorded 
at the \FourS\ resonance, corresponding to $88.84\pm 0.98$ million \BB pairs.
An additional $9.58\,\fb^{-1}$ data sample taken 40 MeV below
the \FourS\ resonance is used in order to study the continuum background 
$e^+ e^- \to q\bar q$ ($q=u,d,s,c$).

Charged particles are detected and their momenta measured with the
combination of a silicon vertex tracker with five layers of
double-sided detectors and a 40-layer central drift chamber, both operating
in the 1.5-T magnetic field of a solenoid.
Charged-particle identification (PID) is provided by measurement of
the average energy loss (dE/dx) in the tracking devices and by an
internally reflecting ring-imaging Cherenkov detector covering the
central region.
A ${K}/\pi$ separation of better than four standard deviations
($\sigma$) is achieved for momenta below 3 GeV/$c$, decreasing to 2.5
$\sigma$ at the highest momenta found in \B decay. Photons and
electrons are detected by a CsI(Tl) electromagnetic calorimeter (EMC).

%% file: method.tex
\section{ANALYSIS METHOD}
\label{sec:method}

We reconstruct \bkstorho candidates through the decays ${K}^{*0} \to K^+ \pi^-$ and 
$\rho^+ \to \pi^+ \pi^0$, $\pi^0 \to \gamma \gamma$.  Inclusion of the charge 
conjugate processes is implied in this paper.

The main backgrounds are from continuum events and ${B}\bar{B}$
events (\B background). Event selection reduces these backgrounds but as a maximum likelihood fit will be performed the selection is kept loose. Some
non-resonant charmless \B decays into four bodies are a particular
problem as they have similar final states to our signal and their
branching fractions are poorly known.

\subsection{Selection and discriminating variables}
\label{sec:selection} 
Monte Carlo (MC) simulations \cite{babar_mc} of the signal and the
\B background as well as the off-resonance beam data were used to
determine the selection criteria before the examination of the
on-resonance beam data.

The final state of the signal decay is $K^+ \pi^+ \pi^- \pi^0$.
We first select charged kaons and pions from charged tracks.
Charged tracks candidates are required to originate from the interaction point: 
distance of closest approach to the interaction point  less than 10 cm along the beam
directions, and less than 1.5 cm in the plane transverse to the beam directions.
We require that the 
charged pion candidates from the ${K}^{*0}$ and $\rho^+$ decays not be identified as
electrons, kaons, or protons, and that the charged kaon candidate from the ${K}^{*0}$ decay 
agrees with a kaon hypothesis and be inconsistent with the electron and proton hypotheses.

Next, we reconstruct $\pi^0$ candidates from photon pairs, where
each photon has an energy larger than 50 MeV and exhibits a lateral
profile of energy deposition in the EMC consistent with an
electromagnetic shower \cite{detector_babar}. The $\pi^0$ candidate
mass must satisfy $0.11<m_{\gamma \gamma}<0.16$ GeV/$c^2$.

We then reconstruct ${K}^{*0}$ and $\rho^+$ candidates.  The mass of
the $K^{\star0}$ and the $\rho^+$ candidates must satisfy
$|m_{{K}^{+}\pi^-} - 0.896| < 0.125$ \gevcc and $|m_{\pi^+ \pi^0} -
0.769| < 0.375$ \gevcc. These mass windows correspond to about 
$ 2.5$ times the resonance full width (Fig. \ref{fig:proj_plots}).
Combinatorial backgrounds dominate near the helicity angle regions
$|\cos \theta _{K^{*0},\rho^+}|=1$, where the vector mesons tend to decay in a
way that produces a soft particle.
The effect is more important when this soft particle is a $\pi^0$.
If one assumes that the longitudinal polarization is large, as
suggested by recent measurements of charmless vector-vector modes and
theoretical predictions, it is important to maintain the largest
possible coverage in $\cos \theta _{K^{*0},\rho^+}$.
The requirements $-0.95 <\cos \theta _{K^{*0}} <1.0$ ($-0.8 <\cos \theta _{\rho^+} <0.95$)
are applied to reject candidates with a soft $\pi^-$ ($\pi^0$ and
$\pi^+$) coming from the $K^{*0}$ ($\rho^{+}$) decay; they allow the
suppression of most of the combinatorial backgrounds while maintaining
adequate efficiency for longitudinally-polarized decays.

We select \B candidates from the $K^{*0}\rho^{+}$ combinations
using two nearly independent kinematic
observables \cite{detector_babar}, the beam energy-substituted \B mass
$\mes \equiv \sqrt{(\frac{s}{2}+ \mathbf{p}_i \mathbf{p}_B)^2/E_i^2 -p_B^2}$
and the  energy difference
$\de \equiv (E_i E_B - \mathbf{p}_i . \mathbf{p}_B - \frac{s}{2})/\sqrt{s}$, 
where $\sqrt{s}$ is the beam energy in the $\Upsilon (4S)$ CM, and
($E_B, \mathbf{p}_B$) and ($E_i, \mathbf{p}_i$) are the
four-momenta of the \B candidate and the $e^+e^-$ initial state,
both defined in the laboratory frame.
For signal events, the $m_{ES}$ distribution peaks at the \B -meson
mass and the $\Delta E$ distribution peaks near zero. We require
\B candidates to satisfy $5.21<m_{ES}<5.29$ GeV/$c^2$ and $|\Delta
E|<0.15$ GeV.  When multiple \B candidates exist in the same event, we
select the one whose reconstructed $\pi^0$ mass is nearest to the
known $\pi^0$ mass; we choose the candidate randomly from those that
share the same \piz.

To discriminate signal from continuum background, we also
use a neural network (NN) combining six variables: a Fisher
discriminant made from two event-shape variables (see \cite{fisher});
the cosine of the angle between the direction of the \B and the
collision axis (z) in the CM frame; the cosine of the angle between
the \B -thrust axis and the z axis; the cosine of the angle between the
\B -thrust axis and the thrust of the particles of the rest of the
event; the angle between the
direction of the $\pi^{0}$ and that of one of its daughter photons in the $\pi^{0}$ rest frame ($\pi^0$
decay angle); 
and the sum of transverse momenta relative to the z-axis of the particles in the
rest of the event.

\subsection{Sample composition}
\label{sec:samp_compo}
The expected numbers of signal and background events in the data
sample are summarized in Table \ref{tab:composition}. The \B background is divided into categories, 
described below, that are modeled separately in the maximum
likelihood fit (Sec. \ref{sec:likelihood}).

\begin{table}[!h]
\caption{Number of events expected in the $K^{*0} \rho ^+$ analysis
 for an 81.85 $fb^{-1}$ data sample.  A branching fraction of $20
 \times 10^{-6}$ and a longitudinal polarization of 75\% are assumed
 for the signal.}
\begin{center}
\begin{tabular}{lc}
Category  &  Expected number \\
& of events    \\
\hline
\hline
Signal                  &  167 \\
\hline
Continuum               &  12400 \\ 
\hline
$b \to charm$  backgrounds  &           \\ 
\hline
$ B^{+}  \rightarrow \bar D^{0} \pi^+$     &  409 \\
$ B^{+}  \rightarrow \bar D^{0} \rho^+$    &  696 \\  
$ B^{+}  \rightarrow c$, excluding $ \bar D^{0} \pi^+$ and $ \bar D^{0} \rho^+$ &   915  \\
$ B^{0}\:\rightarrow c$                    &  904 \\
\hline
$b \to charmless$  backgrounds                  &     \\ 
\hline
$ \rho^{+} \rho ^0$ (100\% longitudinal polarization) & 3.2 \\
$ K^{*0} \pi^0 $                                & 4.3 \\
``Five bodies''                                 & 68  \\
``Other Charmless''                             & 291 \\
\hline
\hline
\end{tabular}
\end{center}
\label{tab:composition}
\end{table}

\subsubsection{Signal}
The signal consists of correctly reconstructed ``true signal'' events
as well as badly reconstructed ``Self-Cross-Feed'' (SxF) events.  SxF
events contain useful information as one of the two vector mesons is
usually correctly reconstructed. They are treated
separately in the fit.

The SxF events are due mostly to the misassignment of a charged (neutral) pion
in about 60\% (40\%) of the cases. The SxF events are
decomposed into three categories: events in which the $K^{*0}$ is
reconstructed correctly but not the $\rho^+$, events in which the
$\rho^+$ is reconstructed correctly but not the $K^{*0}$, and events
in which neither vector meson is reconstructed correctly. 

The total signal selection efficiency, including both ``true signal'' and SxF
events, is $12.6 \pm 0.1\%$ ($20.6 \pm 0.2\%$) for longitudinally (transversely)
polarized events. The fraction of SxF events is $25.1 \pm 0.2 \%$ ($10.9 \pm 0.3\%$) for
longitudinally (transversely) polarized events.

\subsubsection{\B background}
Background from \B decays can be split into that from $b\to c$
transitions and that from $b\to charmless$ transitions. We further
split the $b\to c$ background into four subcategories, each with its
own term in the likelihood fit (Sec. \ref{sec:likelihood}):
$B^+ \to \bar D^0 \pi^+$,
$B^+ \to \bar D^0 \rho^+$,
$B^+ \to charm$ other than the first two, and
$B^0 \to charm$.
The first two peak in $m_{ES}$ while the last two do not.
In particular, $B^+ \to \bar D^0 \pi^+, ~\bar D^0 \to K^+ \pi^- \pi^0$ and 
$B^+ \to \bar D^0 \rho^+, ~\bar D^0 \to K^+ \pi^-$
share the same final state $K^+ \pi^+ \pi^- \pi^0$ 
with signal. 

The other charmless \B backgrounds are modeled in four
categories in the likelihood fit.  The two \brhorhoo and $B^0 \to
K^{*0} \pi^0$ specific modes are separated due to their similarity
with the signal.  The rest of the charmless are divided into two
additional categories.  The ``five bodies'' category includes decay
modes that have five particles in the final state and involve an
intermediate resonance $[a_1^0, a_1^+, a_0^0, a_0^+, \omega, f_0,
\rho^+, \rho^0, K^{*+}, K^{*0}]$ or one of the particles $K^+$, $\pi^+$
or $\pi^0$.  The ``Other Charmless'' category consists of all other
charmless modes after all the above modes have been excluded.  
The modes in the ``five bodies'' category  resemble the
signal more than the ones in the ``Other Charmless'' category due to
the resonances. Therefore their $m_{ES}$ distribution peaks slightly more
 at the \B-meson mass. 
However the  yield of 
the ``Other Charmless'' category is four times larger (Table \ref{tab:composition}).  
Since it is the most  poorly known, it will be floated in the likelihood fit.

None of the non-resonant charmless modes ${K}^{*0} \pi^+ \pi^0$,
${K}^{+} \rho^+ \pi^-$, and ${K}^{+} \pi^+ \pi^- \pi^0$, with the same final state 
as the signal, have been studied experimentally. 
No theoretical model is currently available to predict their branching ratios 
and their decay kinematics. Therefore they are not modeled in the likelihood fit in
this preliminary analysis. Yet they could produce substantial false
signals since their final state is the same as for signal
events. A systematic error will be assigned for this fact (Sec.~\ref{sec:syst_nonres}).

To summarize, we have a total of 13 categories of events: the ``true signal'', three SxF categories, the continuum, 
four categories of charm \B-background, and four categories of charmless \B-background. 
In the fit, we float the sum of the yields in the ``true signal'' and the SxF categories, as well
as the yields in the continuum and the ``Other Charmless'' categories.  
The yields in all the other \B-background categories are fixed.

\subsection{Likelihood fit}
\label{sec:likelihood}
We perform an unbinned, extended maximum likelihood fit on the
selected data sample to extract the signal yield, the longitudinal
polarization $f_L$, and the direct-CP-violating asymmetry \acp. 
Seven observables are used:
% 1
$m_{ES}$,
% 2
$\Delta E$,
% 3
the helicity angles 
$\cos(\theta _{K^{*0}})$ and
% 4
$\cos(\theta _{\rho^+})$,
% 5
the reconstructed masses of the two vectors $m _{K^{*0}}$ and
% 6
$m_{\rho^+}$, 
% 7
and the  neural network output.

All significant correlations are taken into account  in the likelihood
function described below or are covered by systematic errors (Sec. \ref{sec:syst_pdf_model}).

The extended likelihood function is then given by:
\begin{equation}
L =  e^{-N^{\prime}}\prod_{i=1}^{N} 
\left(n_{sig} (f^{true}_{Sig}P_{Sig,i}+ \sum_{SxF_k=1,3} f_{SxF_k} P_{SxF_k, i}) 
+  ~n_{cont} P_{cont,i} +~\sum_{\underset{j}{B~backg.}} n_j P_{j,i}
 \right)\label{eqn:likelihood}
\end{equation}
where $n_{sig}$ and $n_{cont}$ are the numbers of signal (including
SxF) and continuum events that are floated in the fit. The numbers of
events $n_j$ in the \B-background category $j$ are all fixed to their
Monte-Carlo expectations, except in the case of the ``Other
Charmless'' category for which the yield is floated. $N^{\prime}(N)$
are the expected (observed) total numbers of events in the data
sample and $i$ is the event index. $f^{true}_{Sig}$ and
$f_{SxF_{1,3}}$ are the fractions of ``true signal'' and of the three
different SxF categories normalized to the total number of signal
events $n_{sig}$. These fractions are different in the longitudinal
and transverse components and are taken from the simulation.

The normalized $P$ Probability Density Function (PDF) in each
category  is the product of the normalized PDFs of each observable,
except for the continuum for which a joint PDF of the correlated
helicities and vector-meson masses is used.

{\bf PDF Models}

The $m_{ES}$, $\Delta E$, NN, $\cos \theta _{K^{*0}}$, $\cos \theta _{\rho^+}$,
$m_{K^{*0}}$, and $m_{\rho^+}$ PDFs of the ``true signal'' and the \B background 
are taken from the simulation, but the means and widths of
the signal Gaussian PDFs for $m_{ES}$ and $\Delta E$ are corrected to account for
the differences between data and Monte-Carlo observed in a $B^+ \to
\bar D^0 \pi^+$ (with $\bar{D^0} \to K^+ \pi^- \pi^0$) control
samples.

For the signal including the three SxF categories, the distributions
of the vector masses and helicities are modeled using for each vector
meson the distributions of equation (\ref{eqn1}) modified by a
function $a_{K^{*0},\rho^+}(\theta _{K^{*0},\rho^+})$ proportional to the probability for
a signal event to be reconstructed, correctly or not, and multiplied by the
probability $P^{true}_{K^{*0},\rho^+}(\theta _{K^{*0},\rho^+})$
($1-P^{true}_{K^{*0},\rho^+}(\theta _{K^{*0},\rho^+})$) to reconstruct correctly (or
not) the $K^{*0}$ or the $\rho^+$; both functions depend on the
vector-meson helicity angle.  Different mass distributions
$Pm_{K^{*0},\rho^+}^{True,False}(m_{K^{*0},\rho^+})$ of the vector-meson mass $m_{K^{*0},\rho^+}$
are used whether the reconstruction of the meson is correct or not.
For example, for the ``true signal'' longitudinal component we 
have: $a_{K^{*0}}(\theta _{K^{*0}}) a_{\rho^+}(\theta _{\rho^+}) \\ \times P^{true}_{K^{*0}}
(\theta _{K^{*0}}) P^{true}_{{\rho^+}} (\theta _{\rho^+}) \times \cos ^2 (\theta _{K^{*0}}) \cos
^2 (\theta _{\rho^+}) \times Pm_{K^{*0}}^{True}(m_{K^{*0}}) Pm_{\rho^+}^{True}(m_{\rho^+}) $. Another
example is the transverse component of the SxF category with a correctly
reconstructed  $K^{*0}$ described by: $a_{K^{*0}}(\theta _{K^{*0}})
a_{\rho^+}(\theta _{\rho^+}) \times P^{true}_{K^{*0}} (\theta _{K^{*0}}) (1-P^{true}_{\rho^+}
(\theta _{\rho^+})) \times \sin ^2 (\theta _{K^{*0}}) \sin ^2 (\theta _{\rho^+}) \times
Pm_{K^{*0}}^{True}(m_{K^{*0}}) Pm_{\rho^+}^{False}(m_{\rho^+}) $. 
The model supposes that the
reconstruction of the two vector mesons is independent. A
systematic uncertainty for this assumption is included in the total
uncertainty (Sec. \ref{sec:syst_pdf_model}).

%%% CONTINUUM %%%%
In the continuum, the $m_{ES}$ distribution is parametrized by an
ARGUS function \cite{argusf}, the $\Delta E$ distribution by a second order
polynomial, and the neural network output (NN) distribution by the
three-parameter function $(1-NN)^{(a_1+~a_2
(1-NN)+~a_3(1-NN)^2)}$. The shape parameters entering the PDFs of the
$m_{ES}$, $\Delta E$, and NN observables are also floated in the fit:
the statistics of the ``on-resonance'' data is much larger than the
``off-resonance'' data, which is used for cross-checks.  To model the
correlation between the mass and the helicity of each vector meson,
non-parametric PDFs are made from two 2-dimensional mass-helicity
histograms extracted from the $m_{ES}$ side-band ($m_{ES}<5.25$ GeV/$c^2$
and $NN<0.4$), after subtraction of the remaining \B background (less
than 10\%).

%% file: syst_check.tex
\section{SYSTEMATIC UNCERTAINTIES}
\label{sec:sys_error} 

The systematic uncertainties are summarized in Tables~\ref{syst1} and
\ref{syst2}.
Table~\ref{syst1} displays the uncertainties on the efficiency of
signal reconstruction and on the total number of $B \bar B$ pairs in
the data set, each which contributes to a multiplicative uncertainty
on the final branching fraction. The quadratic sum of the
uncertainties in Table~\ref{syst1} is reported in Table~\ref{syst2} as
``Signal reconstruction efficiency, number of \BB pairs.'' It is one
of the largest contributions to the overall systematic uncertainty on
the branching fraction measurement.

\begin{table}
  \caption{Summary of the multiplicative systematic uncertainties
$\Delta B$ on the branching fraction {\BR} associated with the
signal reconstruction efficiency and number of \BB.}
  \begin{center}
  \begin{tabular}{lc}
    Source                        & $\Delta B$ \\
    \hline\hline
    \piz reconstruction           &  8.6 \% \\
    Track reconstruction          &  3.9 \% \\
    PID                           &  1.1 \% \\
    Number of \BB pairs           &  1.1 \% \\
    \hline
    Total                         &  9.6 \% \\
    \hline\hline
  \end{tabular}
  \end{center}
\label{syst1}
\end{table}

\begin{table}
\caption{Summary of the systematic uncertainties on the signal yield,
on the branching fraction, and on the polarization. The asymmetric
uncertainty from the non-resonant charmless backgrounds is presented
separately from the other systematics.}
  \begin{center}
  \begin{tabular}{l|cc|c}
    Source & $N_S$ & \BR  & $f_L$ \\
    \hline\hline
    Signal reconstruction efficiency, number of \BB pairs & --         & 9.6\%       & --         \\
    Ratio of efficiencies for long./trans. polarizations  &$\pm 0.0$   & $\pm$0.0\%  & $\pm$0.006 \\
    ``True signal'' and Self-cross-feed fractions         & $\pm 1.7$  & $\pm$1.2\%  & $\pm$0.001 \\
    PDF shapes                          & $\pm 6.4$  &  $\pm$4.5\%   & $\pm$0.019 \\
    NN shape in off-resonance data      & $\pm 4.9$  &  $\pm$3.5\%   & $\pm$0.027 \\
    No. \B backgrounds                  & $\pm 1.8$  &  $\pm$1.3\%   & $\pm$0.010 \\
     \hline
    Total                               & $\pm 8.4$  & $\pm$11.3\%   & $\pm$0.035 \\
    \hline\hline
    Non-resonant charmless backgrounds    & $-15.7$  &  $-11.1\%$    & $\pm$0.020 \\
    ($K^+ \pi^+ \pi^- \pi^0$ final state) &                &             &       \\  
    \hline\hline
  \end{tabular}
  \end{center}
\label{syst2}
\end{table}

\subsection{Fractions of ``True signal'' and SxF categories, and efficiencies}
Table~\ref{syst2} also gives the systematic uncertainties on the
polarization $f_L$, due to the uncertainty on the ratio of the
selection efficiencies for the longitudinal and transverse polarization
components
(used to compute the effective polarization of the selected signal
events).

Other systematics on the signal yield and the polarization are due to
the uncertainties on the relative fractions $f^{true}_{Sig}$ and
$f_{SxF_{1,2,3}}$ of the ``pure signal'' and of the three SxF
categories, which are taken from the
simulation.  Due to correlations between the $K^*$ and $\rho$
detection and reconstruction, the fraction of events in one category
of signal or SxF is slightly different from the product of the
probabilities $P^{true}_{K^{*0},~\rho^+}(\theta _{K^{*0},~\rho^+})$
(Sec. \ref{sec:likelihood}) to reconstruct correctly or not the
vector meson, averaged over the helicity angles.  The difference
between the results obtained with this set of signal and SxF
fractions and the set of values taken directly from the simulation is
taken as the systematic uncertainty.

\subsection{Uncertainties on PDFs shapes} 
\label{sec:syst_pdfs}
The systematic labeled ``PDF shapes'' in Table~\ref{syst2} is due to
the uncertainty of the PDF shapes for the signal, continuum, and
\B backgrounds.  It includes the error on the correction of the
mean and width of the ``true signal'' $m_{ES}$ and $\Delta E$ gaussian
distributions for the differences observed between the data and the
simulation (Sec. \ref{sec:likelihood}).  It includes also systematics
associated with corrections of biases from the PDF models for the
signal, SxF, and continuum (Sec. \ref{sec:syst_pdf_model}), and
uncertainties due to the limited statistics in determining PDF shapes
(Sec. \ref{sec:syst_pdf_stat}).

\subsubsection{Systematics from the PDF model itself}
\label{sec:syst_pdf_model}
We first checked that the cut $\cos \theta _{\rho^+} > -0.8$ is
sufficiently tight not to leave low momentum $\pi^0$ backgrounds that
are not properly simulated: the results are stable when the analysis is
repeated after a tighter selection on the helicity angle
($\cos \theta_{\rho^+} > -0.5$).

The model for the ``true signal'' and the SxF is tested using fully
simulated signal, divided into sub-samples embedded in toy simulations
of the other backgrounds. Each sample has the same number of
signal events and the same polarization as is fitted in data.  A small
bias of $4.1 \pm 1.4$ events is observed on the signal yield as well as
a $(-1.2 \pm 0.5)$\% bias on the polarization.

In the continuum, the masses and helicity distributions are modeled
by using 2-dimensional mass-helicity histograms extracted from the
$m_{ES}$ side-band (Sec. \ref{sec:likelihood}).  A slight bias of
$-2.2\pm 0.2$ events on the signal yield and of $(-0.9\pm0.1)$\% on the
polarization is associated with this model. This is estimated using toy
Monte-Carlo samples, generated with the PDF used for the data fit. For
each toy experiment new PDFs of the masses and helicities of the
vector mesons are constructed following the same procedure as in the
real data sample. The results obtained when using the new PDFs are
slightly biased compared to the results obtained when using the PDFs
the toy sample was generated from.

The two biases from the signal and continuum models are corrected for
in the final result, and conservative systematic errors, each equal to
half of the correction, are assigned.

\subsubsection{Systematics from the limited statistics to determine the PDFs}
\label{sec:syst_pdf_stat}
The systematic uncertainty associated to the shape of the parametrized
PDFs is determined by varying the parameters within their statistical
error obtained from a fit on the full simulation.  The non-parametric
PDFs are varied by generating toy samples using the original PDFs used
in the real data fit. For each toy sample, new PDFs are made out
of the generated distributions of the observables. The systematic
error is the dispersion of the differences in the fit results between
the new and the original PDFs.

\subsubsection{Neural network shape parameters in the continuum}
\label{sec:syst_pdf_NN}
For the continuum, the $m_{ES}$, $\Delta E$, and NN PDFs are
parametrized. The parameters are determined from real data
(Sec. \ref{sec:likelihood}). Their values agree with the values fitted
on the off-resonance data for the $m_{ES}$ and $\Delta E$
distributions, but not for the NN. A systematic uncertainty is assigned,
labeled ``NN shape in off-resonance data'' in Table~\ref{syst2}. It is
equal to the difference on the results when using the two sets of
parameters.  Though it is an uncertainty on a PDF shape, it is
displayed separately from the ``PDF shapes'' category as it is
relatively large.

\subsection{Number of events in the \B-background categories}  
Other systematic uncertainties, labeled ``No. \B backgrounds'' in
Table~\ref{syst2}, come from the uncertainty on the numbers of events
in each \B-background category that are fixed in the fit. The
branching ratio of the \brhorhoo decay is varied within the errors of
the combined measurements of \cite{babar_vv} and \cite{belle_rhorho}:
$\cal B$(\brhorhoo)$=(26.4^{+6.1}_{-6.4})\times 10^{-6}$. The unknown
branching fraction of the $K^{*0}\pi^0$ background mode is varied by
100\%, assuming a central value of $7.5\times 10^{-6}$. This 
value is  half of the value of the measured branching fraction for
the $K^{*0}\pi^+$  decay mode \cite{babar_kpipi} ($\approx 15\times 10^{-6}$), 
based on the isospin rule and the assumption of penguin dominance. 
The number of
events is varied by 20\% in all the $b \to c$ categories and by 50\%
in the ``Five bodies'' category of charmless-\B background. Note that
there is no systematic error assigned for the ``Other Charmless''
category of events as this yield is floated in the fit to data.

\subsection{Contribution from non-resonant charmless backgrounds decaying into $K^+ \pi^+ \pi^- \pi^0$}
\label{sec:syst_nonres}

The non-resonant charmless \B -decay modes ${K}^{*0} \pi^+ \pi^0$,
${K}^{+} \rho^+ \pi^-$, and ${K}^{+} \pi^+ \pi^- \pi^0$ have the same
final state as the signal and are not modeled in the fit.
The associated systematic error is estimated by the difference in the
data fit result when the yields of these background modes are floated
or fixed to zero: PDFs of these modes can be constructed from their
full simulation, based on a simplified phase space model, but the
available statistics is limited to a few hundreds of events. 
We obtain an asymmetric error of $-15.7$ events ($-11.1$\%) on the signal yield,
which is systematically overestimated when these background modes are
not modeled in the fit. We get an additional symmetric systematic
error of $2.0\%$ on the polarization.  This main systematic error is
a preliminary estimation and is presented separately
from the other systematics, with the label ``non-resonant''. 

Note that this study seems to show that most of these backgrounds are negligible,
probably because their decay
products populate different regions of the phase space. The only channel which has
overlapping phase space is $(K^+ \pi^-)_{S-wave} \rho^+$\!. A more detailed 
study of these backgrounds will be necessary.

\section{PHYSICS RESULTS}
We obtain from the fit $147.3^{+23.4}_{-22.3}\mathrm{(stat)}$ signal events and
a longitudinal polarization fraction
$f_L=0.77 \pm 0.08\mathrm{(stat)}$.
We then correct for the two small biases described in
Section~\ref{sec:syst_pdf_model}, from the models of the signal
and self-cross-feed, and of the continuum.
We obtain
$141.0^{+23.4}_{-22.3}\mathrm{(stat)} \pm {15.9}\mathrm{(syst)} ^{+0.0}_{-15.7}
(\mbox{non-resonant})$
signal events and
 $f_L=0.79\pm 0.08\mathrm{(stat)}\pm 0.04 \mathrm{(syst)} \pm 0.02 (\mbox{non-resonant})$.
The polarization observed is consistent with both the polarization
found in the other pure penguin modes such as $K^* \phi$ ($f_L\sim
0.5$) and with purely longitudinal polarization ($f_L\sim 1$).
From the number of signal events, the fraction of longitudinal
polarization, the selection efficiencies determined for the
transverse and longitudinal polarization components, and  
the branching fractions ${\cal B}(K^{*0}\to K^+ \pi^-)$, 
${\cal B}(\pi^{0}\to \gamma \gamma)$,
we compute the branching fraction:
$${\cal B}(\bkstorho)=
[17.0\pm{}{2.9}\mathrm{(stat)}\pm{}{2.0}\mathrm{(syst)}{}^{+0.0}_{-1.9}(\mbox{non-resonant})]\times 10^{-6}.
$$
The impact of the uncertainties on
$\cal B$($K^{*0}\to K^+ \pi^-$) and
$\cal B$($\pi^{0}\to \gamma \gamma$)
is negligible compared to the other systematic errors.    
\begin{figure}[!h]\begin{center}
  \includegraphics[width=10cm]{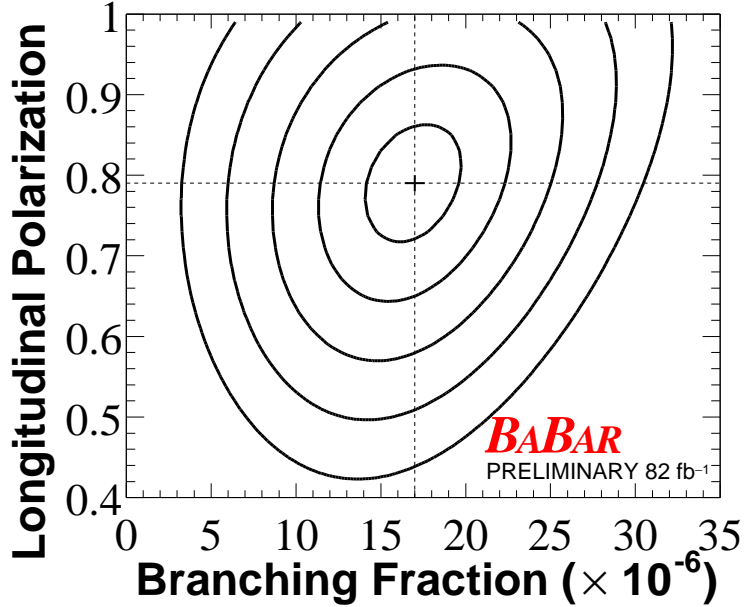}
  \caption{Countours at the 1, 2, 3, 4, and 5 $\sigma$ levels showing the correlated statistical uncertainty on the branching fraction and on the polarization.} 
  \label{fig:ellipse_BR_fl}
  \end{center}\end{figure}
The statistical error on the branching fraction results from the  
statistical errors on the signal yield and the polarization, taking into account their
correlation (Fig. \ref{fig:ellipse_BR_fl}). 
The  systematic uncertainty on the branching fraction results from 
the propagation of the systematics uncertainties  on the signal yield and on the polarization. 
The systematic error on the signal yield (resp. polarization) of 15.9 events (resp. 4\%), results in an 
systematic error of $1.9 \times 10^{-6}$ (resp. $0.4 \times 10^{-6}$) on the branching fraction.
Finally, the ``non-resonant'' systematics on the signal yield is propagated to the branching fraction.

This result is consistent with the isospin rule predicting that
${\cal B}(\bkstorho)\approx 2 \times$ ${\cal B}(\bkstrhoo)$, assuming the dominance of
gluonic penguin diagrams in the \bkstrhoo decay and the measured
\bkstrhoo branching fraction of $[10.6 ^{+3.8}_{-3.5}] \times 10^{-6}$
\cite{babar_vv}. This result is about 2 standard deviations away from the 
predictions from naive factorization models \cite{Ali:1998,Gregory:2004} (Table~\ref{tab:summary_predict_meas}).

We also extract the direct-CP-violating charge asymmetry:
$${\cal A}_{CP}(\bkstorho) = [-14\pm 17\mathrm{(stat)}\pm 4\mathrm{(syst)}]\%.$$
It is consistent with zero, as expected for this pure penguin mode.
The systematic uncertainties on \acp arises from
a limit on the possible size of charge-dependent tracking efficiency and 
particle identification biases. The uncertainty from the tracking is the linear sum of the 0.3\% uncertainty for each charged track.
The uncertainty from the particle identification is estimated at 4\%.

\begin{figure}[!h]\begin{center}
\resizebox{15.5cm}{!}{
        \includegraphics{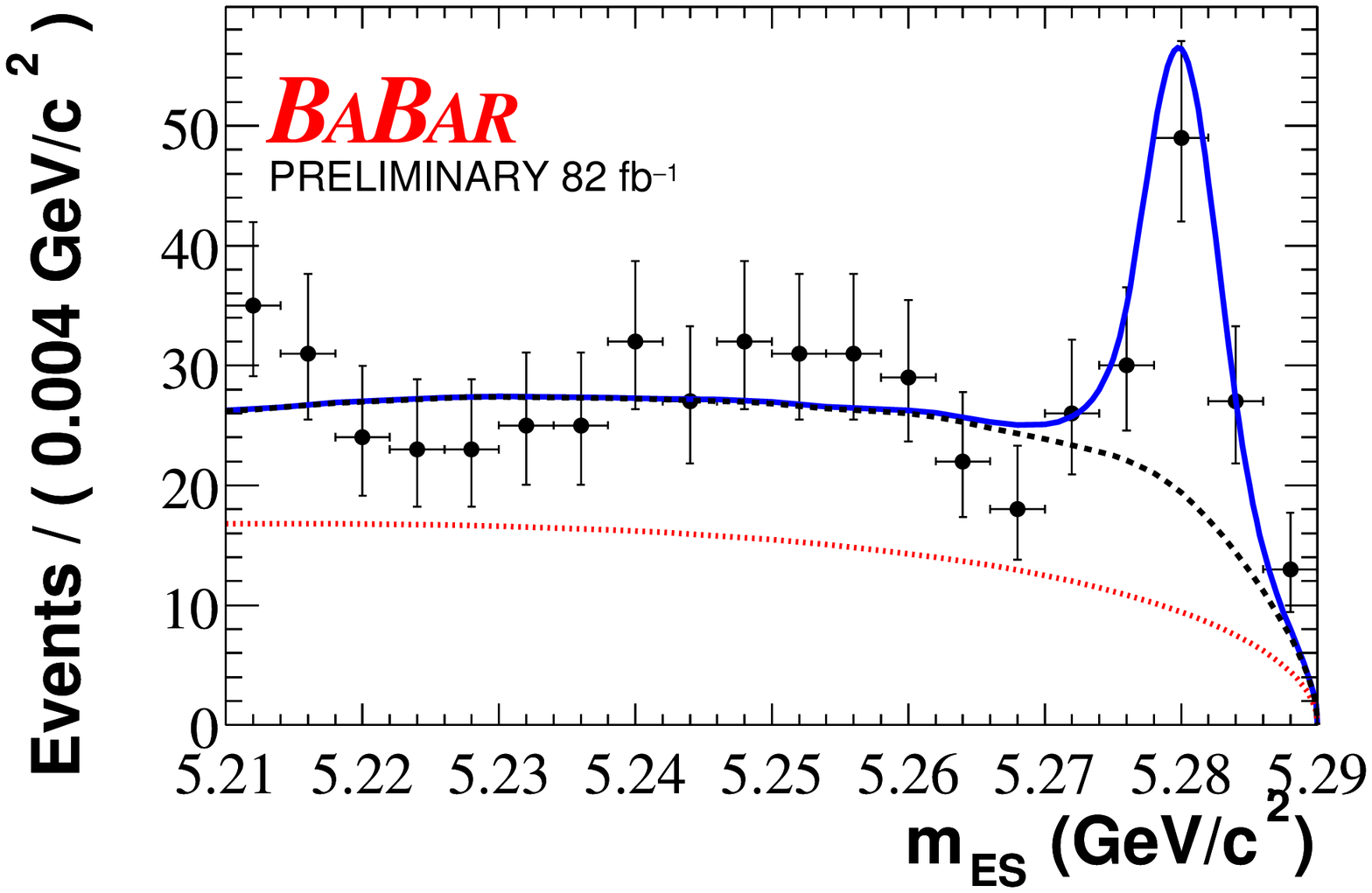}
	\hspace{0.1cm}
        \includegraphics{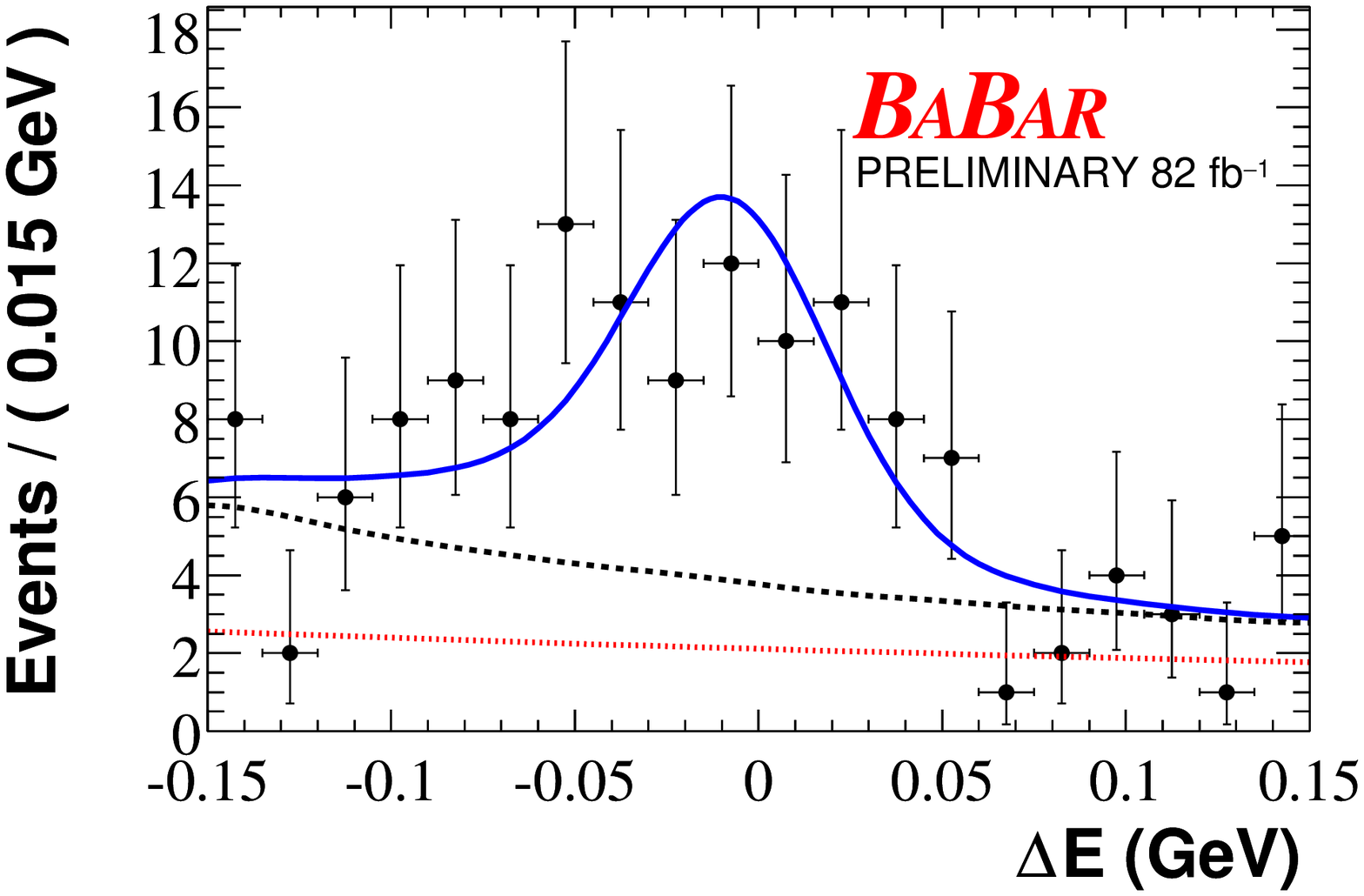}
}
\vspace{0.1cm}
\resizebox{15.5cm}{!}{
        \includegraphics{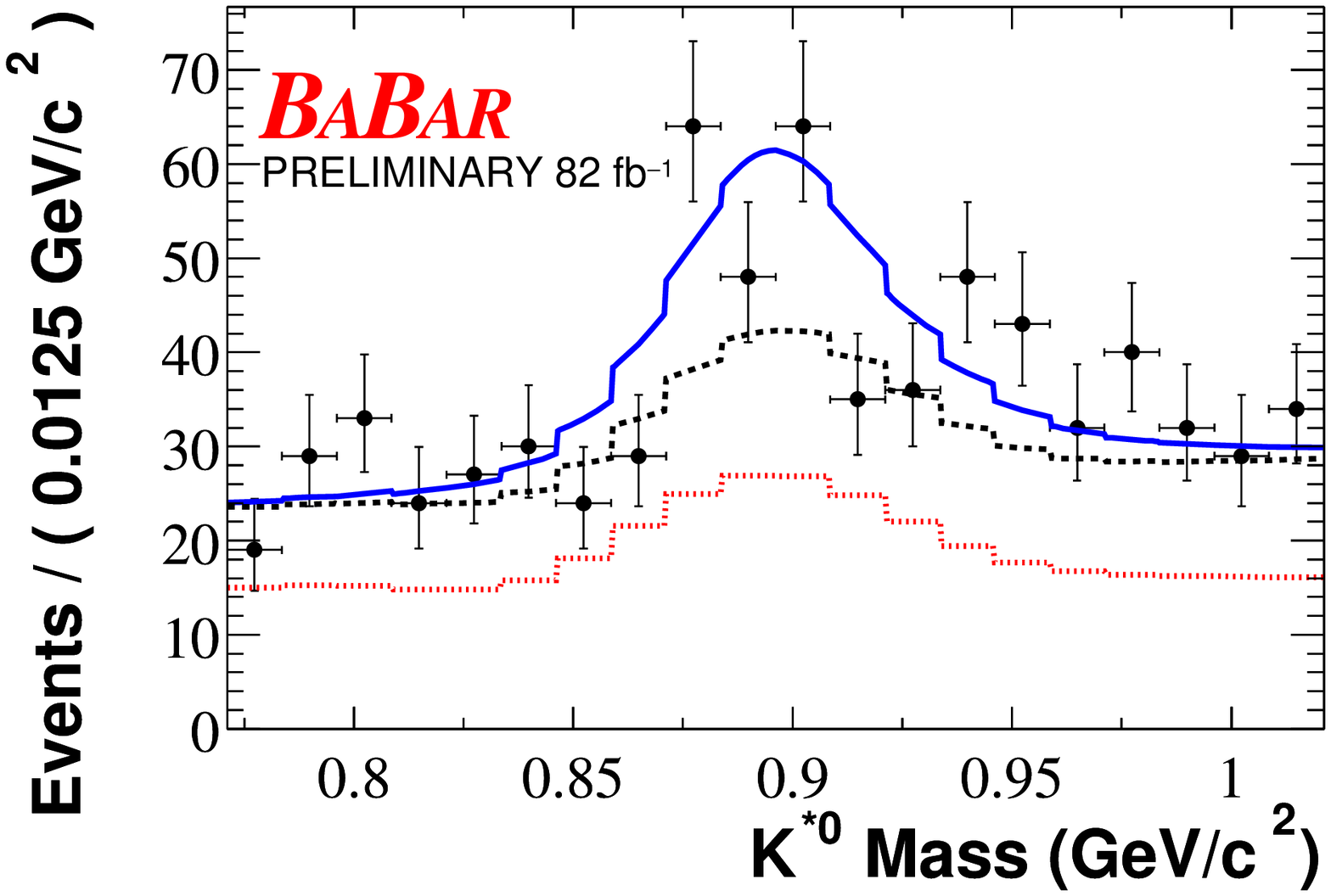}
	\hspace{0.1cm}
        \includegraphics{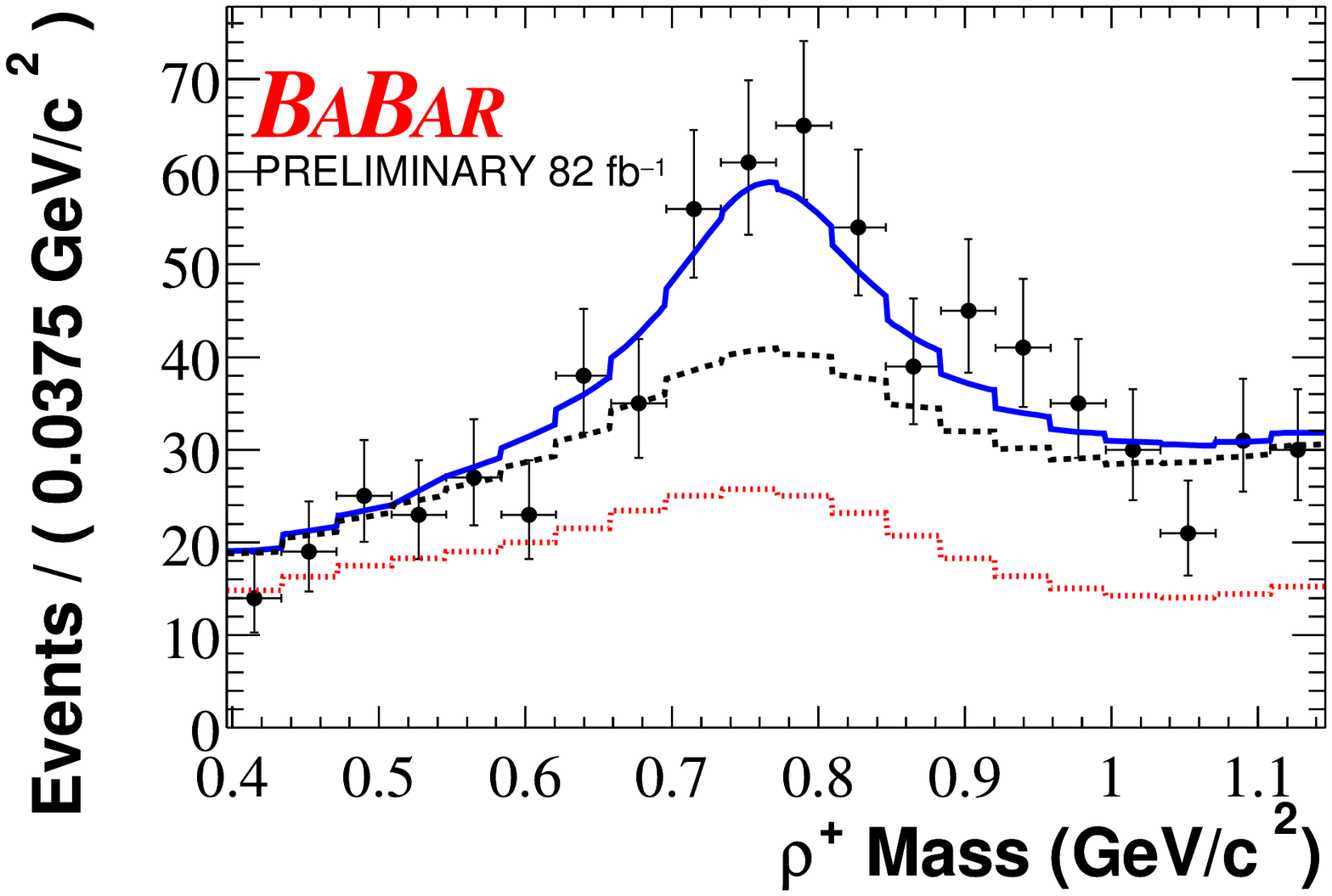}
}
\vspace{0.1cm}
\resizebox{14.0cm}{!}{
\includegraphics{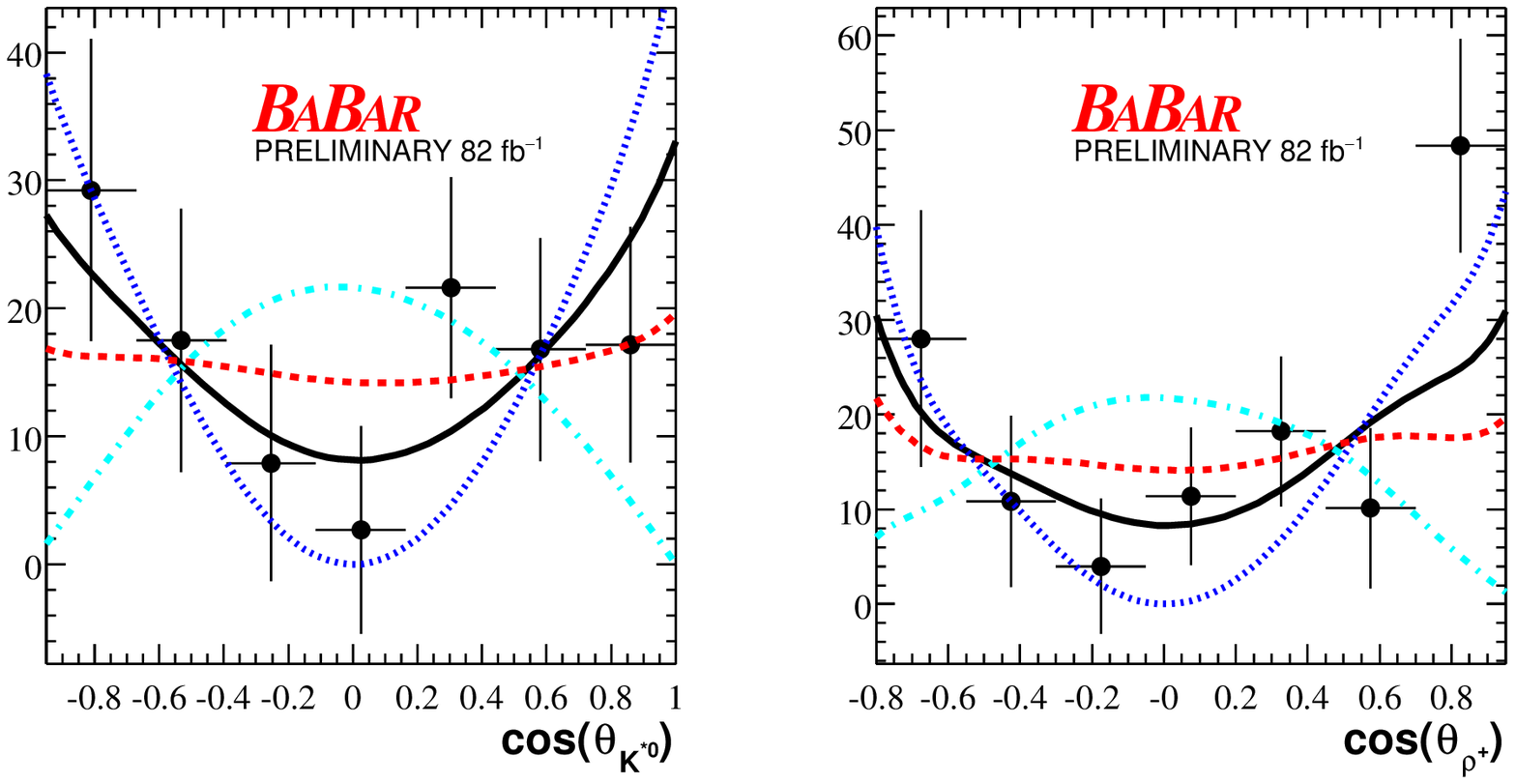}
}
\caption{Distributions of $m_{ES}$, $\Delta E$, $K^{*0}$ and $\rho^+$ 
reconstructed masses and helicities in the \bkstorho decay, enhanced in signal component
by selecting
events with large signal to background likelihood ratios on the
discriminating variables not shown in each plot. 
In the top four plots, the black dots are the data, the blue plain line is the
fitted distribution for the full data sample, the black dashed line is
the fitted distribution for all backgrounds, and the dotted red line
is the distribution of only the continuum events. The two plots in the bottom show 
the distributions of the helicity angles. The data (black dots), after subtraction of the backgrounds,
are compared to the signal component of the fitted PDF (plain black curve). In dashed 
lines are the results expected for different fractions of longitudinal
component: 100\% (blue), 50\% (red), 0\% (cyan). The  pure transverse case is excluded by
the data.}
\label{fig:proj_plots}
\end{center}\end{figure}

The distributions of $m_{ES}$, $\Delta E$, and the vector-meson
helicities and reconstructed masses are shown in
Figure~\ref{fig:proj_plots} with the signal enriched by selecting
events with large signal to background likelihood ratios on the
discriminating variables not shown in each plot.

\section{SUMMARY}
We measure the branching fraction and the fraction of longitudinal
component for the decay \bkstorho using a maximum likelihood
technique. We use a data set corresponding to a total integrated
luminosity of 81.85 fb$^{-1}$ taken on the $\Upsilon(4S)$ peak.  A signal is observed for the first time with a significance of greater than $5\sigma$. From
a fitted signal yield of $141.0^{+23.4}_{-22.3}\mathrm{(stat)} \pm {8.4}\mathrm{(syst)}^{+0.0}_{-15.7}
(\mbox{non-resonant})$ events we obtain the branching fraction:
$$\BR(\bkstorho) =
[17.0\pm{}{2.9}\mathrm{(stat)}\pm{}{2.0}\mathrm{(syst)}{}^{+0.0}_{-1.9}(\mbox{non-resonant})]\times 10^{-6},
$$
the longitudinal polarization fraction $f_L$:
$$f_L=0.79\pm 0.08\mathrm{(stat)}\pm 0.04 \mathrm{(syst)} \pm 0.02 (\mbox{non-resonant}),$$

and the direct-CP-violating asymmetry:
$${\cal A}_{CP}(\bkstorho) = [-14\pm 17\mathrm{(stat)}\pm 4\mathrm{(syst)}]\%.$$
These results are preliminary.

\section{ACKNOWLEDGMENTS}
\input{acknowledgements}

%% file: acknowledgements.tex
We are grateful for the 
extraordinary contributions of our \pep2\ colleagues in
achieving the excellent luminosity and machine conditions
that have made this work possible.
The success of this project also relies critically on the 
expertise and dedication of the computing organizations that 
support \babar.
The collaborating institutions wish to thank 
SLAC for its support and the kind hospitality extended to them. 
This work is supported by the
US Department of Energy
and National Science Foundation, the
Natural Sciences and Engineering Research Council (Canada),
Institute of High Energy Physics (China), the
Commissariat \`a l'Energie Atomique and
Institut National de Physique Nucl\'eaire et de Physique des Particules
(France), the
Bundesministerium f\"ur Bildung und Forschung and
Deutsche Forschungsgemeinschaft
(Germany), the
Istituto Nazionale di Fisica Nucleare (Italy),
the Foundation for Fundamental Research on Matter (The Netherlands),
the Research Council of Norway, the
Ministry of Science and Technology of the Russian Federation, and the
Particle Physics and Astronomy Research Council (United Kingdom). 
Individuals have received support from 
CONACyT (Mexico),
the A. P. Sloan Foundation, 
the Research Corporation,
and the Alexander von Humboldt Foundation.

%% file: main.bbl
\begin{thebibliography}{10}

\bibitem{Fleisher}
A.~Buras, R.~Fleischer, S.~Recksiegel, and F.~Schwab.
\newblock Anatomy of prominent ${B}$ and ${K}$ decays and signatures of
  ${CP}$--violating new physics in the electroweak penguin sector.
\newblock (2004).
\newblock hep-ph/0402112.

\bibitem{NeubertGRL}
M.~Neubert and J.L. Rosner.
\newblock Determination of the weak phase $\gamma$ from rate measurements in
  ${B}^{\pm} \to \pi {K}, \pi \pi$ decays.
\newblock {\em Phys. Rev. Lett.}, {\bf 81}:5076--5079, (1998).
\newblock hep-ph/9809311.

\bibitem{babar_vv}
B.~Aubert \emph{et al.} [BABAR~Collaboration].
\newblock Rates, polarizations, and asymmetries in charmless vector--vector
  ${B}$ meson decays.
\newblock {\em Phys. Rev. Lett.}, {\bf 91}:171802, (2003).
\newblock hep-ex/0307026.

\bibitem{belle_rhorho}
J.~Zhang \emph{et al.} [BELLE~Collaboration].
\newblock Observation of ${B^+}\to\rho^+\rho^0$.
\newblock {\em Phys. Rev. Lett.}, {\bf 91}:221801, (2003).
\newblock hep-ex/0306007.

\bibitem{Wisconsin}
B.~Aubert \emph{et al.} [BABAR~Collaboration].
\newblock Search for the decay ${B}^0 \to {K}^{*+}\rho^-$.
\newblock (2004).
\newblock BABAR-CONF-04/041, SLAC-PUB-10636, hep-ex/0408035.

\bibitem{Ali:1998}
A.~Ali, G.~Kramer, and Cai-Dian Lu.
\newblock Experimental tests of factorization in charmless non-leptonic
  two-body {\B} decays.
\newblock {\em Phys. Rev.}, D{\bf 58}:094009, (1998).
\newblock hep-ph/9804363.

\bibitem{Aleksan}
R.~Aleksan, P.~F. Giraud, V.~Morenas, O.~Pene, and A.~S. Safir.
\newblock Testing ${QCD}$ factorization and charming penguins in charmless
  ${B}\to {P}{V}$.
\newblock {\em Phys. Rev.}, D{\bf 67}:094019, (2003).
\newblock hep-ph/0301165.

\bibitem{kagan_polvv}
A.~Kagan.
\newblock Polarization in ${B} \to {VV}$ decays.
\newblock (2004).
\newblock hep-ph/0405134, submitted to Phys. Lett. B.

\bibitem{babar_rhoprhom}
B.~Aubert \emph{et al.} [BABAR~Collaboration].
\newblock Observation of the decay ${B}^{0}\to \rho^{+}\rho^{-}$ and
  measurement of the branching fraction and polarization.
\newblock {\em Phys. Rev.}, D{\bf 69}:031102, (2004).
\newblock hep-ex/0311017.

\bibitem{babar_phikstar}
B.~Aubert \emph{et al.} [BABAR~Collaboration].
\newblock Measurement of the ${B}^0 \to \phi {K}^{*0}$ decay amplitude.
\newblock (2004).
\newblock BABAR-PUB-04/031.

\bibitem{belle_phikstar}
A.~Bosek \emph{et al.} [BELLE~Collaboration].
\newblock {\em Phys. Rev. Lett.}, {\bf 91}:201801, (2003).
\newblock hep-ex/0307014.

\bibitem{ygrossman}
Y.~Grossman.
\newblock {\em Int. J. Mod. Phys.}, A{\bf 19}:907, (2004).

\bibitem{Colangelo:2004rd}
P.~Colangelo, F.~De~Fazio, and T.~N. Pham.
\newblock The riddle of polarization in ${B} \to {VV}$ transitions.
\newblock (2004).
\newblock hep-ph/0406162.

\bibitem{Gregory:2004}
G.~Schott.
\newblock Study of the rare decays ${B}\to {K}^{*}\rho/\rho\rho$ and search for
  {$CP$} violation in these modes in the {\babar} experiment.
\newblock {\em \emph{Ph.D. Thesis, Paris 6 University}}, July (2004).

\bibitem{formal_vvdecay}
G.~Kramer and W.F. Palmer.
\newblock {\em Phys. Rev.}, D{\bf 45}:193, (1992).

\bibitem{detector_babar}
B.~Aubert \emph{et al.} [BABAR~Collaboration].
\newblock {\em Nucl. Instrum. Methods}, A{\bf 479}:1--116, (2002).

\bibitem{babar_mc}
S.~Agostinelli \emph{et al.} [GEANT4~Collaboration].
\newblock {\em Nucl. Instrum. Methods}, A{\bf 506}:250, (2003).

\bibitem{fisher}
B.~Aubert \emph{et al.} [BABAR~Collaboration].
\newblock {\em Phys. Rev. Lett.}, {\bf 89}:281802, (2002).

\bibitem{argusf}
H.~Albrecht \emph{et al.} [ARGUS~Collaboration].
\newblock {\em Z. Phys.}, C{\bf 48}:543, (1990).

\bibitem{babar_kpipi}
B.~Aubert \emph{et al.} [BABAR~Collaboration].
\newblock Measurements of the branching fractions of charged ${B}$ decays to
  ${K}^{\pm}\pi^{\mp}\pi^{\pm}$ final states.
\newblock (2003).
\newblock hep-ex/0308065, Submitted to Phys. Rev. D.

\end{thebibliography}
